\documentclass[10pt,conference]{IEEEtran}
\usepackage{graphicx}
\usepackage{amssymb,amsmath}

\begin{document}

\title{Waterfilling Theorems in the Time-Frequency Plane for the Heat Channel and a Related Source}

\author{
\IEEEauthorblockN{Edwin Hammerich}
\IEEEauthorblockA{Ministry of Defence\\
Kulmbacher Str. 58--60, D-95030 Hof, Germany\\
E-mail: edwin.hammerich@ieee.org}
}
\maketitle

\begin{abstract}
The capacity of the heat channel, a linear time-varying (LTV) filter with additive white Gaussian noise 
(AWGN), is characterized by waterfilling in the time-frequency plane. Similarly, the rate distortion 
function for a related nonstationary source is characterized by reverse waterfilling in the 
time-frequency plane. The source is formed by the white Gaussian noise response of the same LTV filter as
before. The proofs of both waterfilling theorems rely on a specific Szeg\H{o} theorem for a positive 
definite operator associated with the filter. An essentially self-contained proof of the Szeg\H{o} 
theorem is given. The waterfilling theorems compare well with classical results of Gallager and 
Berger. In case of the nonstationary source it is observed that the part of the classical power spectral 
density (PSD) is taken by the Wigner-Ville spectrum (WVS).
\end{abstract}

\newtheorem{definition}{Definition}
\newtheorem{theorem}{Theorem}
\newtheorem{lemma}{Lemma}
\newtheorem{remark}{Remark}

\section{Introduction}
The characterization of the capacity of continuous-time (or waveform) channels by waterfilling in the 
frequency domain, going back to Shannon \cite{Shannon1949}, has been given by Gallager \cite{Gallager} 
for linear time-invariant (LTI) waveform channels in great generality. At least since the advent of 
mobile communications there is a vivid interest in similar results for LTV channels; see 
\cite{Barbarossa}, \cite{Jung}, \cite{Farrell} to cite only a few. Despite some progress---for instance 
LTV filter descriptions by pseudodifferential operators \cite{Groch} based on 
the Weyl symbol \cite{Kozek}, the spreading function \cite{Farrell} or others \cite{Jung}---no (simple) 
characterizations of the capacity of LTV channels are known. The waterfilling theorems in the 
present paper may provide helpful examples in this connection (and beyond).

We consider the operator $\boldsymbol{P}_\delta^{(\gamma)}$ from the Hilbert space 
$L^2(\mathbb{R})$ of square-integrable functions $f:\mathbb{R}\rightarrow\mathbb{C}\!\cup\!\{\infty\}$ 
into itself given by
\begin{eqnarray}
\lefteqn{(\boldsymbol{P}_\delta^{(\gamma)}f)(t)=e^{-\frac{t^2}{2\alpha^2}}}  \label{TFLO}\\
	&&\cdot\frac{\beta}{\sqrt{2\pi\cosh\delta}}\int_{-\infty}^\infty \exp\left[-\frac{\beta^2}{2}
	                           \left(\frac{t}{\cosh\delta}-t'\right)^2\right]f(t')\,dt',   \nonumber
\end{eqnarray}
where $\alpha,\beta$ are any positive numbers satisfying $\alpha\beta>1$ and $\gamma,\delta>0$ are 
defined by $\gamma^2=\alpha/\beta$, $\coth\delta=\alpha\beta$ \cite{Ham2004}. This operator, in its 
original form introduced as time-frequency localization operator in signal analysis \cite{Daub}, is 
deeply rooted in quantum mechanics (see \cite{Daub}, \cite{Ham2009}, \cite{Janssen} and the 
references therein); suffice it to say that the condition $\alpha\beta>1$ is a manifestation of the 
uncertainty principle. The operator~\eqref{TFLO} regarded as LTV filter for finite-energy  signals 
$f(t)$ will play a central role in our investigations. Then, $t$ is time, $\alpha$ has the physical 
dimension of time and $\beta$ that of a frequency.

\section{The Heat Channel Revisited} \label{Sec_II}
In contrast to \cite{Ham2009}, we define the heat channel as the continuous-time LTV channel
\begin{equation}
   \tilde{g}(t)=(\boldsymbol{P}_\delta^{(\gamma)}f)(t)+n(t),\,-\infty<t<\infty,  \label{HC}
\end{equation}
where $\boldsymbol{P}_\delta^{(\gamma)}$ is the LTV filter \eqref{TFLO}, the real-valued filter input 
signals $f(t)$ are of finite energy and the noise signals  $n(t)$ at the filter output are realizations 
of white Gaussian noise with two-sided PSD $N_0/2=\theta^2>0$.

We now reduce the continuous-time heat channel to a (discrete) vector Gaussian channel following the 
approach in \cite{Gallager} for LTI waveform channels; our analysis is much simplified by 
the restriction to finite-energy input signals. Henceforth, we put 
$\rho=e^{-\delta},\,\delta=\mathrm{arccoth}(\alpha\beta)\sim\frac{1}{\alpha\beta}\,
(\alpha\beta\rightarrow\infty)$ \cite{Ham2009}. For the LTV filter \eqref{TFLO} we have the 
diagonalization \cite{Daub}, \cite{Ham2009}
\begin{equation}
  (\boldsymbol{P}_\delta^{(\gamma)}f)(t)=
                        \sum_{k=0}^\infty\rho^{k+\frac{1}{2}}a_k\,(D_\gamma H_k)(t), \label{EXP}
\end{equation}
where $(D_\gamma H_k)(t)=\gamma^{-\frac{1}{2}}H_k(t/\gamma)$ is the dilated $k$th Hermite 
function $H_k(t)$ and the coefficients are $a_k=\langle f,D_\gamma H_k \rangle$, 
$\langle f_1,f_2 \rangle=\int_{-\infty}^\infty f_1(t)\overline{f_2(t)}\,dt$ denoting the inner product 
in $L^2(\mathbb{R})$. $\{D_\gamma H_k;k=0,1,\ldots\}$ forms a complete orthonormal system in 
$L^2(\mathbb{R})$. The perturbed filter output signal 
$g=\boldsymbol{P}_\delta^{(\gamma)}f$, $\tilde{g}(t)=g(t)+n(t)$, is passed through a bank of matched 
filters with impulse responses $h_k(t)=(D_\gamma H_k)(-t),\,k=0,1,\ldots.$ The matched filter output 
signals are sampled at time zero to yield $\langle\tilde{g}(t),h_k(-t)\rangle=b_k+n_k$, where 
$b_k=\langle g(t),h_k(-t)\rangle=\rho^{k+\frac{1}{2}}a_k$, and the detection errors 
$n_k=\langle n(t),h_k(-t)\rangle$ are realizations of independent identically distributed zero-mean 
Gaussian random variables $N_k$ with the variance $\theta^2$, $N_k\sim\mathcal{N}(0,\theta^2)$. From the 
detected values $\hat{b}_k=b_k+n_k$ we get the estimates 
$\hat{a}_k=\rho^{-k-\frac{1}{2}}\hat{b}_k=a_k+z_k$ for the coefficients $a_k$ of the input 
signal $f$, where $z_k$ are realizations of independent Gaussian random variables 
$Z_k\sim\mathcal{N}(0,\theta^2\rho^{-2k-1})$. Thus, we are led to the infinite-dimensional 
vector Gaussian channel
\begin{equation}
  Y_k=X_k+Z_k,\,k=0,1,\ldots,  \label{HC_discr}
\end{equation}
where the noise $Z_k$ is distributed as described; it is the same channel as in 
\cite[Def.~1]{Ham2009}. Notice that the noise PSD $\theta^2$, measured in watts/Hz, has also 
the dimension of an energy.

The interpretation of the ubiquitous time-frequency product $\alpha\beta$ as degrees of freedom (DoF) 
of filter output signals (cf. \cite{Ham2009}) will be further substantiated in Section~\ref{Sec_V_A}.

\section{A Specific Szeg\H{o} Theorem} \label{Sec_III}
For a linear operator $\boldsymbol{A}:L^2(\mathbb{R})\rightarrow L^2(\mathbb{R})$ the Weyl symbol 
$\sigma_{\boldsymbol{A}}(x,\xi)$---when existing \cite{Groch}---is defined by \cite{Kozek}, 
\cite{Janssen}
\[
(\boldsymbol{A}f)(x)=\frac{1}{2\pi}\iint_{\mathbb{R}^2}\sigma_{\boldsymbol{A}}
                         \left(\frac{x+y}{2},\xi\right)e^{i(x-y)\xi}f(y)\,dy\,d\xi.
\]
The linear map $\boldsymbol{A}\mapsto\sigma_{\boldsymbol{A}}(x,\xi)$ (or its inverse) is called 
Weyl correspondence. For example, the positive definite operator 
$\boldsymbol{A}=(\boldsymbol{P}_\delta^{(\gamma)})^*\boldsymbol{P}_\delta^{(\gamma)}
=\boldsymbol{P}_{2\delta}^{(\gamma)}$ has the Weyl symbol \cite{Ham2004}
\begin{align}
   \sigma_{\boldsymbol{A}}(x,\xi)&=\frac{1}{\cosh\delta}
                             e^{-(\tanh\delta)(\gamma^{-2}x^2+\gamma^2\xi^2)}    \label{WS1}\\
   &=\frac{1}{\cosh\delta}\exp\left(-\frac{x^2}{\alpha^2}-\frac{\xi^2}{\beta^2}\right).\label{WS2}
\end{align}

From now on, $\boldsymbol{A}$ will always stand for the foregoing operator and we shall write 
$\lambda_k\triangleq\rho^{2k+1},\,k=0,1,\ldots,$ for its eigenvalues. The proof of the subsequent 
Szeg\H{o} theorem, Thm.~\ref{SzegoTh}, is inspired by \cite{Janssen} although the Szeg\H{o} 
theorems in \cite{Janssen} are inadequate for our purposes.

\begin{definition} \label{def_2} For any two functions $A,\,B:(1,\infty)\rightarrow\mathbb{R}$ 
the notation $A\doteq B$ means
\[
  \lim_{x\rightarrow\infty}\frac{A(x)-B(x)}{x}=0,                                        \label{doteq}
\]
or, equivalently, $A(x)=B(x)+o(x)$ as $x\rightarrow\infty$, where $o(\cdot)$ denotes the standard Landau 
little-o symbol.
\end{definition}
In our context, $x$ will always be $\alpha\beta$. Thus $A\doteq B$ implies 
that $A(\alpha\beta)/(\alpha\beta)=B(\alpha\beta)/(\alpha\beta)+\epsilon$ where $\epsilon\rightarrow 0$ 
as $\alpha\beta\rightarrow\infty$.
\begin{lemma}\label{Lemma} For any polynomial $G_N(x,z)=\sum_{n=1}^N c_n(x) z^n$ 
with bounded variable coefficients $c_n(x)\in\mathbb{R}$, $x\in(1,\infty),$ it holds
\[
  \sum_{k=0}^\infty G_N(\alpha\beta,\lambda_k)\doteq\frac{1}{2\pi}
         \iint_{\mathbb{R}^2}G_N\left(\alpha\beta,\sigma_{\boldsymbol{A}}(x,\xi)\right)
	                                                                                   \,dx\,d\xi.
\]
\end{lemma}
\begin{IEEEproof}
First, representation (\ref{EXP}) (substitute $\delta$ by $2\delta$ and $\rho$ by $\rho^2$) yields for 
any $f\in L^2(\mathbb{R})$ the expansion
\[G_N(\alpha\beta,\boldsymbol{A})f=\sum_{k=0}^\infty G_N(\alpha\beta,\lambda_k)\langle f,
                                                              D_\gamma H_k \rangle D_\gamma H_k.
\]
Hence, operator $\boldsymbol{B}\triangleq G_N(\alpha\beta,\boldsymbol{A})$ has the trace
\begin{equation}
  \mathrm{tr}\,\boldsymbol{B}=\sum_{k=0}^\infty G_N(\alpha\beta,\lambda_k),   \label{trace_1}
\end{equation}
the series being converging since $G_N(x,0)=0\,\forall x\in(1,\infty)$.

Second, we use the key observation \cite[\emph{trace rule} (0.4)]{Janssen} to obtain (here and 
thereafter, double integrals extend over $\mathbb{R}^2$)
\[
  \mathrm{tr}\,\boldsymbol{B}=\frac{1}{2\pi}\iint\sigma_{\boldsymbol{B}}(x,\xi)\,dx\,d\xi,
\]
where $\sigma_{\boldsymbol{B}}(x,\xi)$ is the Weyl symbol of operator $\boldsymbol{B}$. By linearity of 
the Weyl correspondence, $\sigma_{\boldsymbol{B}}(x,\xi)$ has the expansion 
\[
   \sigma_{\boldsymbol{B}}(x,\xi)=\sum_{n=1}^N c_n(\alpha\beta)\sigma_{\boldsymbol{A}^n}(x,\xi).
\]
Since for any $\gamma>0$ held constant the family of operators 
$\{\boldsymbol{P}_\delta^{(\gamma)};\delta>0\}$ forms a semigroup with respect to $\delta$ (see 
\cite{Ham2004}), it follows that $\boldsymbol{A}^n=\boldsymbol{P}_{2n\delta}^{(\gamma)}$. In 
Eq.~\eqref{WS1}, substitute operator $\boldsymbol{A}$ by $\boldsymbol{A}^n$ and parameter $\delta$ by 
$n\delta$. Because of $\tanh(n\delta)=(n\tanh\delta)(1+o(1))$ we then obtain
\begin{eqnarray*}
  \sigma_{\boldsymbol{A}^n}(x,\xi)&=&\frac{1}{\cosh(n\delta)}e^{-\tanh(n\delta)
                                                               (\gamma^{-2}x^2+\gamma^2\xi^2)}\\
        &=&(1+o(1))(\sigma_{\boldsymbol{A}}(x,\xi))^n\\
	& &\cdot\exp\left[-o(1)\left(\frac{x^2}{\alpha^2}+\frac{\xi^2}{\beta^2}\right)\right],
\end{eqnarray*}
where the Landau symbol $o(1)$ stands for various quantities vanishing as $\delta\rightarrow0$ (or 
$\alpha\beta\rightarrow\infty$). We now estimate
\begin{eqnarray}
  \mathrm{tr}\,\boldsymbol{B}&=&\frac{1}{2\pi}\iint \sigma_{\boldsymbol{B}}(x,\xi)
                                                                                  \,dx\,d\xi\nonumber\\
    &=&\left[\frac{1}{2\pi}\iint G_N\left(\alpha\beta,
                      \sigma_{\boldsymbol{A}}(\alpha x',\beta\xi')\right)\,dx'd\xi' +\epsilon\right]
		                                                          \alpha\beta    \nonumber\\
    &=&\frac{1}{2\pi}\iint G_N(\alpha\beta,\sigma_{\boldsymbol{A}}(x,\xi))\,dx\,d\xi+\epsilon\,
                                                                          \alpha\beta,   \label{trace_2}
\end{eqnarray}
where $\epsilon\rightarrow 0$ as $\alpha\beta\rightarrow\infty$. Eq.~\eqref{trace_2} in combination 
with Eq.~\eqref{trace_1} concludes the proof.
\end{IEEEproof}
\begin{theorem}[Szeg\H{o} Theorem]\label{SzegoTh} Let $g:[0,\Delta]\rightarrow\mathbb{R}$, 
$\Delta\in(0,\infty)$, be a continuous function such that $\lim_{x\rightarrow 0+}g(x)/x$ exists. For 
any functions $a,\,b:(1,\infty)\rightarrow\mathbb{R}$, where $a(x)$ is bounded and $b(x)\in[0,\Delta]$, 
define the function $G(x,z)=a(x)g(b(x)z),\,(x,z)\in(1,\infty)\times[0,1]$. Then it holds
\begin{equation}
  \sum_{k=0}^\infty G(\alpha\beta,\lambda_k)\doteq\frac{1}{2\pi}
          \iint_{\mathbb{R}^2}G\left(\alpha\beta,\sigma_{\boldsymbol{A}}(x,\xi)\right)
	                                                                     \,dx\,d\xi.\label{Szego}
\end{equation}
\end{theorem}
\begin{IEEEproof}
The function $f(x)=g(x)/x,\,x\in(0,\Delta],$ has a continuous extension $F(x)$ onto 
the compact interval $[0,\Delta]$. By virtue of the Weierstrass approximation theorem, for 
any $m\in\mathbb{N}$ there exists a polynomial $F_{N_m-1}(x)$ of some degree $N_m-1$ such that 
$|F(x)-F_{N_m-1}(x)|\le\epsilon_m=\frac{1}{m}$ for all $x\in [0,\Delta]$. Consequently, the polynomial 
$g_{N_m}(x)=xF_{N_m-1}(x)$ of degree $N_m$ satisfies the inequality
\begin{equation}
  |g(x)-g_{N_m}(x)|\le \epsilon_m x,\,x\in[0,\Delta]. \label{WAS_ineq}
\end{equation}

Define the polynomial with variable coefficients $G_{N_m}(x,z)=a(x)g_{N_m}(b(x)z)$. We now show that
\begin{equation}
  (\alpha\beta)^{-1}\sum_{k=0}^\infty G_{N_m}(\alpha\beta,\lambda_k)\rightarrow
        (\alpha\beta)^{-1}\sum_{k=0}^\infty G(\alpha\beta,\lambda_k) \label{first_arrow}
\end{equation}
and
\begin{eqnarray}
\lefteqn{\frac{(\alpha\beta)^{-1}}{2\pi}\iint_{\mathbb{R}^2}G_{N_m}\left(\alpha\beta,
                                     \sigma_{\boldsymbol{A}}(x,\xi)\right)\,dx\,d\xi}\nonumber\\
        &\rightarrow&\frac{(\alpha\beta)^{-1}}{2\pi}\iint_{\mathbb{R}^2}G\left(\alpha\beta,
	                             \sigma_{\boldsymbol{A}}(x,\xi)\right)\,dx\,d\xi \label{second_arrow}
\end{eqnarray}
as $m\rightarrow\infty$, uniformly for all $\alpha\beta\in(1,\infty)$.

\textit{Proof of (\ref{first_arrow}):} By Ineq. (\ref{WAS_ineq}) we get
\begin{eqnarray*}
\lefteqn{|\sum_{k=0}^\infty G(\alpha\beta,\lambda_k)-\sum_{k=0}^\infty G_{N_m}(\alpha\beta,\lambda_k)|}\\
     &\le& \sum_{k=0}^\infty|G(\alpha\beta,\lambda_k)-G_{N_m}(\alpha\beta,\lambda_k)|\\
     &\le& M\epsilon_m\Delta\sum_{k=0}^\infty \lambda_k,
\end{eqnarray*}
where $M=\sup\{|a(x)|;x>1\}<\infty$ and $\sum_{k=0}^\infty\lambda_k=\rho/(1-\rho^2)
=\alpha\beta/(2\cosh\delta)<\alpha\beta/2$, $\alpha\beta>1$. After devision of the inequalitiy by 
$\alpha\beta$, convergence in (\ref{first_arrow}) follows as claimed.

\textit{Proof of (\ref{second_arrow}):} Similarly,
\begin{eqnarray*}
\lefteqn{|\iint G\left(\alpha\beta,\sigma_{\boldsymbol{A}}(x,\xi)\right)\,dx\,d\xi}\\
     &&-\iint G_{N_m}\left(\alpha\beta,\sigma_{\boldsymbol{A}}(x,\xi)\right)\,dx\,d\xi|\\
     &\le& M\epsilon_m\Delta\iint\sigma_{\boldsymbol{A}}(x,\xi)\,dx\,d\xi.
\end{eqnarray*}
Since $(2\pi)^{-1}\iint\sigma_{\boldsymbol{A}}(x,\xi)\,dx\,d\xi=\alpha\beta/(2\cosh\delta)$, 
after division by $2\pi\alpha\beta$ we come to the same conclusion as before.

Now, choose a (large) number $m\in\mathbb{N}$, so that the left-hand sides in \eqref{first_arrow}, 
\eqref{second_arrow} become arbitrarily close to their respective limit. Replace function $G$ in 
Eq.~\eqref{Szego} with the polynomial $G_{N_m}$. Then, by Lem.~\ref{Lemma} and the uniform convergence 
in \eqref{first_arrow}, \eqref{second_arrow}, the theorem follows. 
\end{IEEEproof}

\section{Waterfilling Theorem for the Capacity of the Heat Channel} \label{Sec_IV}						       
The function $N(t,\omega)$ occurring in the next theorem is defined as
\begin{equation}
    N(t,\omega)=\frac{\theta^2}{2\pi}\cdot(\cosh\delta)
                  \exp\left(\frac{t^2}{\alpha^2}+\frac{\omega^2}{\beta^2}\right).       \label{N}
\end{equation}
$O(\cdot)$ denotes the standard Landau big-O symbol and $x^+$ the positive part of $x\in\mathbb{R}$, 
$x^+=\max\{0,x\}$.
\begin{theorem}\label{WFT1} Assume that the average energy $S$ of the input signal depends on 
$\alpha\beta$ such that $S(\alpha\beta)=O(\alpha\beta)$ as $\alpha\beta\rightarrow\infty$. 
Then for the capacity (in nats per transmission) of the heat channel it holds
\begin{equation}
   C\doteq \frac{1}{2\pi}\iint_{\mathbb{R}^2}\frac{1}{2}
       \ln\left(1+\frac{(\nu-N(t,\omega))^+}{N(t,\omega)}\right)\,dt\,d\omega,   \label{C}
\end{equation}
where $\nu$ is chosen so that
\begin{equation}
   S\doteq\iint_{\mathbb{R}^2}(\nu-N(t,\omega))^+\,dt\,d\omega.   \label{S}
\end{equation}
\end{theorem}
\begin{IEEEproof}
The first part of the proof is accomplished by waterfilling on the noise variances 
\cite[Thm.~7.5.1]{Gallager} (as in the proof of \cite[Thm.~1]{Ham2009}). Let 
$\nu_k^2=\theta^2\rho^{-2k-1}=\theta^2\lambda_k^{-1},\,k=0,1,\ldots,$ be the noise variance in the $k$th 
subchannel of the discretized heat channel \eqref{HC_discr}. The ``water level"  $\sigma^2>\nu_0^2$ 
(excluding the trivial case $S=0$) is defined by the condition
\begin{equation}
  S = \sum_{k=0}^\infty (\sigma^2-\nu_k^2)^+=\sum_{k=0}^{K-1} (\sigma^2-\nu_k^2),\label{def_sigma}
\end{equation}
where $K=\max\{k\in\mathbb{N};\nu_{k-1}^2<\sigma^2\}$ 
is the number of subchannels in the resulting finite-dimensional vector Gaussian channel. The capacity 
$C$ of that vector channel is achieved when the components $X_k$ of the input vector 
$(X_0,\ldots,X_{K-1})$ are independent random variables $\sim \mathcal{N}(0,\sigma^2-\nu_k^2)$; then
\begin{equation}
  C=\sum_{k=0}^{K-1}\frac{1}{2}\ln\left(1+\frac{\sigma^2-\nu_k^2}{\nu_k^2}\right)
                                                                      \quad\mathrm{nats}. \label{C1}   
\end{equation}

In the second part of the proof we apply the above Szeg\H{o} theorem, Thm.~\ref{SzegoTh}. To start 
with, note that $\sigma^2$ is dependent on $\alpha\beta$ and that always 
$\sigma^2=\sigma^2(\alpha\beta)>\theta^2$. On the other hand, the function $\sigma^2(\alpha\beta)$ is 
finitely upper bounded as $\alpha\beta\rightarrow\infty$ because of the growth condition imposed on 
$S=S(\alpha\beta)$ and the equation (refer to the proof of \cite[Thm.~1]{Ham2009})
\[ 
  S\doteq\frac{\alpha\beta}{2}\theta^2\cdot\left(\frac{\sigma^2}{\theta^2}\ln\frac{\sigma^2}{\theta^2}
                                                -\frac{\sigma^2}{\theta^2}+1\right),
\]
observing that the function $y=x\ln x-x+1$ is positive and convex downward for $x>1$. Define 
\begin{equation}
  \ln_+ x=\left\{\!\!\begin{array}{cl}\max\{0,\ln x\} & \mbox{if }x>0,\\
                                                    0 & \mbox{if }x=0.
                     \end{array}\right.                                        \label{ln+}
\end{equation}
By Eq.~\eqref{C1} we have
\begin{align*}
  C&=\sum_{k=0}^\infty\frac{1}{2}\ln_+\left(\frac{\sigma^2(\alpha\beta)}{\theta^2}
                                                                                     \lambda_k\right)\\
   &=\sum_{k=0}^\infty a(\alpha\beta)g(b(\alpha\beta)\lambda_k),
\end{align*}
where $a(\alpha\beta)=1$, $b(\alpha\beta)=\sigma^2(\alpha\beta)/\theta^2$, 
$g(x)=\frac{1}{2}\ln_+x,x\in[0,\Delta]$, and $\Delta$ is chosen so that $b(\alpha\beta)\le\Delta<\infty$ 
when $\alpha\beta$ is large enough. The latter choice is possible since $\sigma^2(\alpha\beta)$ remains 
bounded as $\alpha\beta\rightarrow\infty$; without loss of generality ($\mathrm{w.l.o.g.}$), we assume 
$b(\alpha\beta)\in[0,\Delta]$ for \emph{all} $\alpha\beta\in(1,\infty)$. Then, by Thm.~\ref{SzegoTh} it 
follows that $C=C(\alpha\beta)$ satisfies
\begin{align*}
    C&\doteq\frac{1}{2\pi}\iint
       \frac{1}{2}\ln_+\left(\frac{\sigma^2(\alpha\beta)}{\theta^2}\sigma_{\boldsymbol{A}}(x,\xi)\right)
                                                                                             \,dx\,d\xi\\
     &=\frac{1}{2\pi}\iint\frac{1}{2}
 \ln\!\left[1+\frac{\left(\frac{\sigma^2(\alpha\beta)}{2\pi}-N(t,\omega)\right)^+}{N(t,\omega)}
                                                                                   \right]\!dt\,d\omega,
\end{align*}
where $N(t,\omega)\triangleq\frac{\theta^2}{2\pi}\sigma_{\boldsymbol{A}}(t,\omega)^{-1}$. Next, rewrite 
Eq.~\eqref{def_sigma} as
\[
   S=\sum_{k=0}^\infty\sigma^2(\alpha\beta)\left(1
                            -\frac{1}{\frac{\sigma^2(\alpha\beta)}{\theta^2}\lambda_k}\right)^+.
\]
Put $a(\alpha\beta)=\sigma^2(\alpha\beta)$, $b(\alpha\beta)=\sigma^2(\alpha\beta)/\theta^2$ and define
\[
    g(x)=\left\{\!\!\begin{array}{cl}\left(1-\frac{1}{x}\right)^+ & \mbox{if }x>0,\\
                                                                0 & \mbox{if }x=0.
                     \end{array}\right.
\] 
Again, $\mathrm{w.l.o.g.}$, we may assume that $a(\alpha\beta)$ is bounded and 
$b(\alpha\beta)\in[0,\Delta]$ for all $\alpha\beta\in(1,\infty)$ where $\Delta$ is chosen as above. 
Then, by Thm.~\ref{SzegoTh} it follows that
\begin{align*}
   S&\doteq\frac{1}{2\pi}\iint\sigma^2(\alpha\beta)\left(1-
      \frac{1}{\frac{\sigma^2(\alpha\beta)}{\theta^2}\sigma_{\boldsymbol{A}}(x,\xi)}\right)^+\,dx\,d\xi\\
    &=\iint \left(\frac{\sigma^2(\alpha\beta)}{2\pi}-N(t,\omega)\right)^+\,dt\,d\omega.
\end{align*}

Finally, replacement of $\frac{\sigma^2(\alpha\beta)}{2\pi}$ by the parameter $\nu$ completes the proof.
\end{IEEEproof}
\begin{remark}
Note that the use of Landau symbols does not mean that we need to pass to the limit 
(here as $\alpha\beta\rightarrow\infty$). Rather, the dotted equations \eqref{C},~\eqref{S} may give 
useful approximations even when $\alpha\beta$ is finite (but large enough). 
\end{remark}
For example, if $\alpha\beta=50$ and $S=20$, $\theta^2=0.01$ (units omitted), then Eq.~\eqref{C} yields 
with $\nu=0.051$ an approximate capacity of 75.1043 nats/transmission---the exact one (determined 
numerically) is 75.1017. For that $\nu$, the integral on the right-hand side of Eq.~\eqref{S} 
evaluates to 20.0013---a value close to $S$. Actually, $\nu$ has been computed as described in the above 
proof.

Eqs.~\eqref{C},~\eqref{S} may also be taken for a parametric representation of the function $C=C(S)$ 
(neglecting the error terms). To get rid of the latter one might prefer to average with respect to the 
DoF $\alpha\beta$ and letting $\alpha\beta\rightarrow\infty$.

When $\beta$ is held constant and $\alpha\rightarrow\infty$, the LTV channel \eqref{HC} appears to tend 
towards an LTI waveform channel according to Gallager's model in \cite[Ch.~8]{Gallager} with LTI filter 
with impulse response $h_1(t)=(\beta/\sqrt{2\pi})\exp(-\beta^2t^2/2)$ (we stick to the notations in 
\cite{Gallager}). It is therefore interesting to compare Thm.~\ref{WFT1} with Gallager's capacity 
theorem \cite[Thm.~8.5.1]{Gallager} when applied to that particular waveform channel (with AWGN of 
noise PSD $N_0/2=\theta^2$). By Gallager's theorem we obtain for the capacity $C$ (in bits per 
second) and average input \emph{power} $S$ the parametric representation
\begin{align}
   C&=\frac{1}{2\pi}\int_{-\infty}^\infty\frac{1}{2}
            \log_2\left(1+\frac{(\nu-N_1(\omega))^+}{N_1(\omega)}\right)\,d\omega \label{C_Gall}\\
   S&=\int_{-\infty}^\infty(\nu-N_1(\omega))^+\,d\omega, \label{S_Gall} 	    
\end{align}
where $\nu$ is the parameter, $\omega$ is angular frequency, and
\begin{equation*}
   N_1(\omega)=\frac{\theta^2}{2\pi}\cdot \exp\left(\frac{\omega^2}{\beta^2}\right).
\end{equation*}
We observe perfect formal analogy between the waterfilling formulas \eqref{C_Gall},~\eqref{S_Gall} and 
those in Thm.~\ref{WFT1}. Moreover, $N(t,\omega)$ in \eqref{N} tends to $N_1(\omega)$ as 
$\alpha\rightarrow\infty$ for any $t,\omega$ held constant.

\section{Reverse Waterfilling Theorem for a Related Nonstationary Source} \label{Sec_V}
In the present section we consider the nonstationary source formed by the nonstationary zero-mean 
Gaussian process given by the Karhunen-Lo\`{e}ve expansion
\begin{equation}
  X(t)=\sum_{k=0}^\infty X_k\,(D_\gamma H_k)(t),\,t\in\mathbb{R},\label{KL}
\end{equation}
where the coefficients $X_k,\,k=0,1,\ldots,$ are independent random variables 
$\sim\mathcal{N}(0,\sigma_k^2)$ with the variances 
$\sigma_k^2=\sigma^2\rho^{2k+1}=\sigma^2\lambda_k,\,\sigma>0$. It is the response of the LTV 
filter~\eqref{TFLO} on white Gaussian noise with PSD $N_0/2=\sigma^2$; cf.~\cite{Ham2014}.

\subsection{Wigner-Ville Spectrum of the Source} \label{Sec_V_A}
The WVS $\Phi(t,\omega)$ of the nonstationary random process 
$\{X(t),t\in\mathbb{R}\}$ in \eqref{KL} describes its density of (average) energy in the time-frequency 
plane \cite{Flandrin}. The WVS may be regarded as the nonstationary counterpart to the PSD of a stationary 
random process. It is defined by means of the Wigner transform \cite{Groch} of the realizations 
$x(t)$ of $\{X(t)\}$ and then taking the expectation (for details refer to \cite{Ham2014}). Then 
$ \Phi(t,\omega)=(2\pi)^{-1}\int e^{-i\omega t'}r(t+t'/2,t-t'/2)\,dt'$, where 
$r(t_1,t_2)=\mathbb{E}[X(t_1)\overline{X(t_2)}]$ is the autocorrelation function; in our case we 
obtain \cite{Ham2014}
\begin{equation}
  \Phi(t,\omega)=\frac{\sigma^2}{2\pi}\cdot\frac{1}{\cosh\delta}
                      \exp\left(-\frac{t^2}{\alpha^2}-\frac{\omega^2}{\beta^2}\right). \label{WVS}
\end{equation}
Indeed, the average energy $E =E(\alpha\beta)$ of the process \eqref{KL} is
\begin{equation}
  E =\sum_{k=0}^\infty\sigma_k^2
    =\iint_{\mathbb{R}^2}\Phi(t,\omega)\,dt\,d\omega
    \left(\doteq\frac{\alpha\beta}{2}\sigma^2\right).                                          \label{E}
\end{equation}

Notice that the WVS \eqref{WVS} is proportional to the Weyl symbol \eqref{WS2} (with 
$x\gets t,\,\xi\gets\omega$). 
Consequently, it defines the same ellipse of concentration as described in \cite{Ham2009}.

\subsection{$R(D)$ by Reverse Waterfilling in the Time-Frequency Plane} \label{Sec_V_B}
Substitute the continuous-time Gaussian process $\{X(t),\,t\in\mathbb{R}\}$ in (\ref{KL}) by the
sequence of coefficient random variables $\boldsymbol{X}=X_0,X_1,\ldots\,$. For an estimate 
$\boldsymbol{\hat{X}}=\hat{X}_0,\hat{X}_1,\ldots$ of $\boldsymbol{X}$ we take the squared-error 
distortion $D=\mathbb{E}[\sum_{k=0}^\infty(X_k-\hat{X}_k)^2]$ as distortion measure.

The Landau symbol $\Omega(\cdot)$ occurring in the next theorem is defined for any two functions as in 
Def.~\ref{def_2} as follows: $A(x)=\Omega(B(x))$ as $x\rightarrow\infty$ if $B(x)>0$ and 
$\liminf_{x\rightarrow\infty}A(x)/B(x)>0$.
\begin{theorem}\label{WFT2} Assume that the foregoing average distortion $D$ depends on 
$\alpha\beta$ such that $D(\alpha\beta)=\Omega(\alpha\beta)$ as $\alpha\beta\rightarrow\infty$. Then 
the rate distortion function $R = R(D)$ for the nonstationary source (\ref{KL}) is given by
\begin{equation}
  R\doteq\frac{1}{2\pi}\iint_{\mathbb{R}^2}\max\left\{0,\frac{1}{2}
             \ln\frac{\Phi(t,\omega)}{\lambda}\right\}
	                                                \,dt\,d\omega,   \label{R}
\end{equation}
where $\lambda$ is chosen so that
\begin{equation}
   D\doteq\iint_{\mathbb{R}^2}
             \min\left\{\lambda,\Phi(t,\omega)\right\}\,dt\,d\omega.          \label{D}
\end{equation}
The rate is measured in nats per realization of the source.
\end{theorem}
\begin{IEEEproof}
First, assume $0<D<E$ where $E$ is the average energy \eqref{E}. 
The reverse waterfilling argument for a finite number of independent Gaussian sources \cite{Berger}, 
\cite{Cover} carries over to our case without changes resulting in a finite collection of Gaussian 
sources $X_0,\ldots,X_{K-1}$ where $K=\max\{k\in\mathbb{N};\sigma_{k-1}^2>\theta^2\}$ and the 
``(ground-)water table" $\theta^2>0$ is determined by the condition
\begin{equation}
  D=\sum_{k=0}^\infty \min\{\theta^2,\sigma_k^2\}. \label{D_def}
\end{equation}
Then $K\ge1$ and the rate distortion function $R=R(D)$ for the parallel Gaussian source 
$(X_0,\ldots,X_{K-1})$ is given by \cite[Thm.~10.3.3]{Cover}
\begin{equation}
  R = \sum_{k=0}^{K-1}\frac{1}{2}\ln\frac{\sigma_k^2}{\theta^2}\quad\mathrm{nats}. \label{R_def}
\end{equation}

Now we apply the above Szeg\H{o} theorem, Thm.~\ref{SzegoTh}. Note first that $\theta^2$ is dependent 
on $\alpha\beta$ and that always $\theta^2=\theta^2(\alpha\beta)<\sigma^2$. On the other hand, the 
function $\theta^2(\alpha\beta)$ is positively lower bounded as $\alpha\beta\rightarrow\infty$ because 
of the growth condition imposed on $D=D(\alpha\beta)$ and the equation (refer to the proof of 
\cite[Thm.~2]{Ham2009})
\[
  D\doteq\frac{\alpha\beta}{2}\sigma^2\cdot\left(\frac{\theta^2}{\sigma^2}
                       -\frac{\theta^2}{\sigma^2}\ln \frac{\theta^2}{\sigma^2}\right),
\]
observing that the function $y=x-x\ln x$ is positive and strictly monotonically increasing for $0<x\le 1$ 
and $y\rightarrow 0$ as $x\rightarrow 0+$. By Eq.~\eqref{D_def} we have,
\begin{align*}
  D&=\sum_{k=0}^\infty\theta^2(\alpha\beta)\min\left\{1,\frac{\sigma^2}{\theta^2(\alpha\beta)}
                                                                                  \lambda_k\right\}\\
   &=\sum_{k=0}^\infty a(\alpha\beta)g(b(\alpha\beta)\lambda_k),
\end{align*}
where $a(\alpha\beta)=\theta^2(\alpha\beta)$, $b(\alpha\beta)=\sigma^2/\,\theta^2(\alpha\beta)$, 
$g(x)=\min\{1,x\},\,x\in[0,\Delta]$, and $\Delta$ is chosen so that $b(\alpha\beta)\le\Delta<\infty$ 
when $\alpha\beta$ is large enough. The latter choice is possible since $\theta^2(\alpha\beta)$ remains 
positively lower bounded as $\alpha\beta\rightarrow\infty$; $\mathrm{w.l.o.g.}$ we assume here and 
thereafter that $b(\alpha\beta)\in[0,\Delta]$ for \emph{all} $\alpha\beta\in(1,\infty)$. 
$a(\alpha\beta)$ is always bounded. Therefore, by Thm.~\ref{SzegoTh} we infer
\begin{align*}
  D&\doteq\frac{1}{2\pi}\iint\theta^2(\alpha\beta)
    \min\left\{1,\frac{\sigma^2}{\theta^2(\alpha\beta)}\sigma_{\boldsymbol{A}}(x,\xi)\right\}\,dx\,d\xi\\
   &=\iint\min\left\{\frac{\theta^2(\alpha\beta)}{2\pi},
                                                    \Phi(t,\omega)\right\}\,dt\,d\omega,
\end{align*}
where $\Phi(t,\omega)=\frac{\sigma^2}{2\pi}\,\sigma_{\boldsymbol{A}}(t,\omega)$ is the 
WVS~(\ref{WVS}) of the source. Next, rewrite Eq.~\eqref{R_def} as
\[
  R=\sum_{k=0}^\infty\frac{1}{2}\ln_+\left(\frac{\sigma^2}{\theta^2(\alpha\beta)}\lambda_k\right),
\]
where $\ln_+$ is as defined in \eqref{ln+}. Taking $a(\alpha\beta)=1$, 
$b(\alpha\beta)=\sigma^2/\,\theta^2(\alpha\beta)$, $g(x)=\frac{1}{2}\ln_+x,x\in[0,\Delta]$, $\Delta$ 
chosen as before, by Thm.~\ref{SzegoTh} it follows that
\begin{align*}
  R&\doteq\frac{1}{2\pi}\iint
       \frac{1}{2}\ln_+\left(\frac{\sigma^2}{\theta^2(\alpha\beta)}\sigma_{\boldsymbol{A}}(x,\xi)\right)
                                                                                           \,dx\,d\xi \\
   &=\frac{1}{2\pi}\iint\frac{1}{2}
       \ln_+\left[\frac{\Phi(t,\omega)}{\frac{\theta^2(\alpha\beta)}{2\pi}}
                                                                                 \right]\,dt\,d\omega.
\end{align*}

Finally, replacement of $\frac{\theta^2(\alpha\beta)}{2\pi}$ by the parameter $\lambda$ concludes the 
proof in case $0<D<E$.

When $D=E$ (the case $D=0$ is precluded by asssumption as 
$\alpha\beta\rightarrow\infty$; the case $D>E$ is of no interest) then, as always, $R(D)=0$. Choosing 
in the theorem the parameter 
$\lambda=\Phi(0,0)=\max_{t,\omega}\Phi(t,\omega)$, we obtain correctly 
$R\doteq0,\,D\doteq E$. This completes the proof of the theorem.
\end{IEEEproof}

As in Section~\ref{Sec_IV}, Eqs.~\eqref{R},~\eqref{D} may also be taken for a \emph{parametric} 
representation of the $R(D)$ function. In parametric form, the $R(D)$ function has 
been given by Berger \cite{Berger} for a broad class of stationary random processes. In the latter 
parametric interpretation, Eq.~\eqref{R} is in perfect analogy to \cite[Eq.~(4.5.52)]{Berger}
(with WVS instead of PSD), likewise Eq.~\eqref{D} with regard to \cite[Eq.~(4.5.51)]{Berger} (apart 
from a factor $\frac{1}{2\pi}$).

\section{Conclusion}
Two waterfilling theorems in the time-frequency plane were stated in terms of the Weyl symbol (or its 
reciprocal) and rigorous proofs have been given. The relevance of the Weyl symbol was reflected by its 
equivalence with the WVS of the nonstationary source. The proof of a specific Szeg\H{o} theorem took 
advantage of the semigroup property of the LTV filter. Although the latter feature is not necessarily a 
prerequisite for a conceivable generalization of the present results, asymptotic analysis and notation 
seem unavoidable.

\end{document}